\documentclass[letterpaper,twocolumn,english,pra,aps,showpacs,floatfix,groupedaddress,superscriptaddress,tightenlines]{revtex4}
\usepackage[]{fontenc}
\usepackage[latin9]{inputenc}
\usepackage{amsthm}
\usepackage{amsmath}
\usepackage{amssymb}
\usepackage{esint}

\makeatletter
\@ifundefined{textcolor}{}
{%
 \definecolor{BLACK}{gray}{0}
 \definecolor{WHITE}{gray}{1}
 \definecolor{RED}{rgb}{1,0,0}
 \definecolor{GREEN}{rgb}{0,1,0}
 \definecolor{BLUE}{rgb}{0,0,1}
 \definecolor{CYAN}{cmyk}{1,0,0,0}
 \definecolor{MAGENTA}{cmyk}{0,1,0,0}
 \definecolor{YELLOW}{cmyk}{0,0,1,0}
 }
\theoremstyle{plain}
\newtheorem{thm}{Theorem}

\newcommand{\hc}{\mathrm{h.c.}}

\newcommand{\1}{\leavevmode{\rm 1\ifmmode\mkern  -4.8mu\else\kern -.3em\fi I}}

\usepackage{times}

\makeatother

\usepackage{babel}

\begin{document}

\title{Equivalence between XY and dimerized models}

\author{Lorenzo Campos Venuti}

\affiliation{Institute for Scientific Interchange, ISI Foundation, Viale S. Severo
65, I-10133 Torino, Italy}

\author{Marco Roncaglia}

\affiliation{Dipartimento di Matematica e Informatica, Università degli Studi
di Salerno, Via Ponte don Melillo, I-84084 Fisciano (SA), Italy}

\affiliation{Institute for Scientific Interchange, ISI Foundation, Viale S. Severo
65, I-10133 Torino, Italy}
\begin{abstract}
The spin-1/2 chain with XY anisotropic coupling in the plane and the
XX isotropic dimerized chain are shown to be equivalent in the bulk.
For finite systems we prove that the equivalence is exact in given
parity sectors, after taking care of the precise boundary conditions.
The proof is given constructively by finding unitary transformations
that map the models onto each other. Moreover, we considerably generalized
our mapping and showed that even in case of fully site dependent couplings
the XY chain can be mapped onto an XX model. This result has potential
application in the study of disordered systems. 
\end{abstract}


\maketitle
Exactly solvable models play an important role as limiting cases of
more complex system or for testing numerical algorithms. Moreover
their physical properties can generally be calculated exactly and
traced back to simple mechanism that can be used in more complicated
scenarios. In this article we consider two notable solvable models:
the anisotropic XY model --originally introduced in \cite{LSM61}
with the aim of gaining insights on the long range properties of the
Heisenberg model-- and the dimerized XX model --used sometimes as
a prototype model to describe spin-Peierls distortion--. We prove
the equivalence of these two models, despite in the literature they
are generally considered as separate. The equivalence is shown directly
by means of a unitary transformation for their fermionic counterpart
and traced back to the spin models, carefully taking care of the boundary
conditions.

\paragraph*{Introduction}

For a chain of length $L$ the dimerized XX and anisotropic XY models
are given by the following Hamiltonians\begin{align}
H_{d}^{\eta} & =\frac{1}{2}\sum_{i=1}^{L}\left(1+\gamma\left(-1\right)^{i}\right)\left[\sigma_{i}^{x}\sigma_{i+1}^{x}+\sigma_{i}^{y}\sigma_{i+1}^{y}\right]\label{eq:H_d-spin}\\
H_{XY}^{\eta} & =\frac{1}{2}\sum_{i=1}^{L}\left[\left(1+\gamma\right)\sigma_{i}^{x}\sigma_{i+1}^{x}+\left(1-\gamma\right)\sigma_{i}^{y}\sigma_{i+1}^{y}\right].\label{eq:H_XY-spin}\end{align}
The superscript $\eta$ denotes different kind of boundary conditions
(BCs), $\sigma_{L+1}^{x,y}=\eta\sigma_{1}^{x,y}$, $\eta=1,-1,0$,
corresponding to periodic (PBC), antiperiodic (ABC), and open (OBC)
respectively. Since both Hamiltonians Eqs.~(\ref{eq:H_d-spin}) and
(\ref{eq:H_XY-spin}) commute with the parity operator $P=\prod_{i}\sigma_{i}^{z}$,
we define the parity sectors $\sigma=\pm1$ and the corresponding
projection operators $\Pi_{\sigma}=\left(\1+\sigma P\right)/2$. The
central result of this article establishes that $H_{d}$ and $H_{XY}$
are unitarily equivalent ($\cong$) up to at most a border term as
precisely stated by the following 
\begin{thm}
For $L$ odd and OBC the models (\ref{eq:H_d-spin}) and (\ref{eq:H_XY-spin})
are unitarily equivalent. For $L$ even and PBC or ABC the equivalence
holds in given parity blocks depending on the boundary conditions,
according to the relation: \[
\Pi_{\sigma}H_{d}^{\eta}\Pi_{\sigma}\cong\Pi_{\eta\left(-1\right)^{L/2}}H_{XY}^{\sigma\left(-1\right)^{L/2}}\Pi_{\eta\left(-1\right)^{L/2}}\]
\label{thm:spin}
\end{thm}
In other words, for $L$ even, the boundary index in one model sets
the parity sector in the other (times a modulation factor $\left(-1\right)^{L/2}$),
i.e.~$\sigma_{XY}=\left(-1\right)^{L/2}\eta_{d}$ and $\sigma_{d}=\left(-1\right)^{L/2}\eta_{XY}$.
An immediate consequence of this result is that the two models share
the same thermodynamics, since for $L\to\infty$ the effect of boundary
terms disappear. 

The reason for considering OBC is partly due to the possibility of
using models Eqs.~(\ref{eq:H_d-spin}) and (\ref{eq:H_XY-spin})
to implement quantum information devices. It has been shown in \cite{LDE}
that in the ground state of the dimer model (though with OBC and $L$
\emph{even}), the end spins tend to entangle considerably already
for small values of the dimerization $\gamma$. Moreover the entanglement
survives in the infinite length limit (long distance entanglement).
In a similar fashion, it was already observed in \cite{LSM61} that
the end-spins of the anisotropic model order and such order survives
in the thermodynamic limit (TDL). However this kind of order is of
classical nature and no entanglement is present between the end-spins
of the open anisotropic chain %
\footnote{For instance, for $\gamma>0$ the only non-zero correlation surviving
in the TDL is $\langle\sigma_{1}^{x}\sigma_{L}^{x}\rangle$ \cite{LSM61}.
Such order is clearly classical.%
}.

Before proceeding to examine the proof of the Theorem, let us spend
few words on some benefits of such result. First, let us note that
both models commute with $\pi$-rotations around axis $x$ and $y$,
$\mathcal{R}_{\pi}^{\alpha}=\prod_{i}e^{i\pi\sigma_{i}^{\alpha}/2}$,
$\alpha=x,\, y$. However, the dimer model $H_{d}$ manifests a much
larger symmetry, the total magnetization $M^{z}=\sum_{i}\sigma_{i}^{z}$.
This means that $H_{d}$ is block diagonal in sectors with given magnetization
$M^{z}$, a feature which is especially useful in case of non-integrable
extensions of $H_{d}$ (which maintain this symmetry) where one has
to resort to numerical diagonalization. Thanks to Theorem \ref{thm:spin},
such a symmetry (or an approximate one) must exist also for the anisotropic
model $H_{XY}$. As we will see the magnetization in the dimer model
is mapped onto a non-local operator which we are able to compute.
Clearly this operator has the same spectrum of $M^{z}$ and commutes
with $H_{XY}$. 

The proof of Theorem \ref{thm:spin} relies on a similar theorem holding
for the fermionic version of the models (denoted here with a tilde),
\begin{align}
\tilde{H}_{d}^{\epsilon} & =\sum_{i=1}^{L}\left(1+\gamma\left(-1\right)^{i}\right)\left[d_{i}^{\dagger}d_{i+1}+d_{i+1}^{\dagger}d_{i}\right]\label{eq:H_d-fermi}\\
\tilde{H}_{XY}^{\epsilon} & =\sum_{i=1}^{L}\left[a_{i}^{\dagger}a_{i+1}+\gamma a_{i}^{\dagger}a_{i+1}^{\dagger}\right]+\hc.\label{eq:H_XY-fermi}\end{align}
Here $\epsilon=1,-1,0$ distinguishes among PBC, ABC, and OBC for
the fermions, i.e. $d_{L+1}=\epsilon d_{1}$ and $a_{L+1}=\epsilon a_{1}$.
As we will see later, the spin systems (\ref{eq:H_d-spin}) and (\ref{eq:H_XY-spin})
are connected to the quadratic fermionic models $\tilde{H}_{d}^{\epsilon}$
and $\tilde{H}_{XY}^{\epsilon}$ via a Jordan-Wigner (JW) transformation,
after careful reshuffling of the boundary conditions. The result for
the fermionic models is
\begin{thm}
In the following cases: $L$ even and PBC or ABC, $L$ odd and OBC,
the models (\ref{eq:H_d-fermi}) and (\ref{eq:H_XY-fermi}) are unitarily
equivalent, i.e.~there exists a unitary operator $U$ (a {}``mapping'')
such that $U\tilde{H}_{d}^{\epsilon}U^{\dagger}=\tilde{H}_{XY}^{\epsilon}$.\label{thm:fermi} 
\end{thm}
A simple way to remind the different cases in which the theorem applies
is given by the following argument. Sending $a_{j}\to ia_{j}$ in
$\tilde{H}_{XY}^{\epsilon}$, one realizes that the spectrum of $\tilde{H}_{XY}^{\epsilon}$
is invariant under the transformation $\gamma\to-\gamma$. By relabeling
the sites of the dimer model, one sees that $\tilde{H}_{d}^{\epsilon}$
possess the same invariance only when it contains an even number of
bonds. This occurs for $L$ even in case of PBC or ABC, while for
$L$ odd only in case of OBC.

\paragraph*{Proof of Theorem 2}

Since the fermionic Hamiltonians are quadratic, one way of proving
the equivalence between them is to show that they have the same one-body
spectrum. To diagonalize the anisotropic model we rewrite the Hamiltonians
following the conventions of \cite{LSM61}: $\tilde{H}_{XY}^{\epsilon}=\sum_{i,j}a_{i}^{\dagger}A_{i,j}a_{j}+\left(1/2\right)\left[\sum_{i,j}a_{i}^{\dagger}B_{i,j}a_{j}^{\dagger}+\hc\right]$
and $\tilde{H}_{d}^{\epsilon}=\sum_{i,j}d_{i}^{\dagger}M_{ij}d_{j}$
with matrices given by\[
A=\left(\begin{array}{cccc}
0 & 1 &  & \epsilon\\
1 & 0 & \ddots\\
 & \ddots & \ddots & 1\\
\epsilon &  & 1 & 0\end{array}\right),\, B=\gamma\left(\begin{array}{cccc}
0 & 1 &  & \epsilon\\
-1 & 0 & \ddots\\
 & \ddots & \ddots & 1\\
-\epsilon &  & -1 & 0\end{array}\right),\]
while $M$ is \[
\left(\begin{array}{cccc}
0 & 1-\gamma & \cdots & \epsilon[1+\left(-1\right)^{L}\gamma]\\
1-\gamma & 0 & \ddots & \vdots\\
\vdots & \ddots & \ddots & 1-\left(-1\right)^{L}\gamma\\
\epsilon[1+\left(-1\right)^{L}\gamma] &  & 1-\left(-1\right)^{L}\gamma & 0\end{array}\right).\]
The one particle energies of $\tilde{H}_{XY}^{\epsilon}$ are given
by the (positive) square root of the eigenvalues of $\left(A-B\right)\left(A+B\right)$.
Calling $\Lambda_{k}$ such roots, since $A$ is traceless, one arrives
at \cite{LSM61} \[
\tilde{H}_{XY}^{\epsilon}=\sum_{k}\Lambda_{k}\eta_{k}^{\dagger}\eta_{k}-\frac{1}{2}\sum_{k}\Lambda_{k}.\]
The equivalence of the two models now stems from the fact that, for
$L$ even and PBC or ABC, and for $L$ odd and OBC $M^{2}=\left(A-B\right)\left(A+B\right)$.
Moreover, under the same hypothesis, the eigenvalues of $M$ are symmetric
around zero (for $L$ odd and OBC there is one zero eigenvalue). To
write $\tilde{H}_{d}^{\epsilon}$ in the same form as $\tilde{H}_{XY}^{\epsilon}$
perform a particle-hole transformation on the negative eigenvalues
of $M$. We arrive then at $\tilde{H}_{d}^{\epsilon}=\sum_{k}\Lambda_{k}\beta_{k}^{\dagger}\beta_{k}-\sum_{\mathrm{neg}}\Lambda_{k}$,
where $\sum_{\mathrm{neg}}$ is the sum over the negative eigenvalues
of $M$. To complete the proof note that, in the specified cases,
$\sum_{\mathrm{neg}}\Lambda_{k}=\left(1/2\right)\sum_{k}\Lambda_{k}$.
$\square$

\paragraph*{The mapping}

The above proof does not give the explicit form of the mapping. We
will now provide a physically more compelling proof which has the
additional advantage of revealing an exact form of the mapping. For
simplicity we will stick to $L$ even and PBC/ABC for the fermionic
models. The first step is to write both models in Fourier space\begin{align}
\tilde{H}_{d}^{\epsilon} & =\sum_{k}\left[2\cos(k)d_{k}^{\dagger}d_{k}+2i\gamma\sin\left(k\right)d_{k+\pi}^{\dagger}d_{k}\right]\label{eq:H_d-Fourier}\\
\tilde{H}_{XY}^{\epsilon} & =\sum_{k}\left\{ 2\cos(k)a_{k}^{\dagger}a_{k}\right.\nonumber \\
 & \left.+\gamma\left[i\sin\left(k\right)a_{k}^{\dagger}a_{-k}^{\dagger}-i\sin\left(k\right)a_{-k}a_{k}\right]\right\} \label{eq:H_XY-Fourier}\end{align}
Let us consider first PBC. The momenta in the Brillouine zone (BZ)
are given by $k=2\pi n/L,$ $n=-L/2+1,\ldots,L/2$. Note that only
for PBC and ABC if $k\in\mathrm{BZ}$ then $-k\in\mathrm{BZ}$. Moreover
only for $L$ even $k\in\mathrm{BZ}\Rightarrow k+\pi\in\mathrm{BZ}$.
In particular Eqs.~(\ref{eq:H_d-Fourier}) and (\ref{eq:H_XY-Fourier})
are not correct if $L$ is odd. The unitary transformation that maps
the dimer model onto the XY is \begin{equation}
d_{k}^{\dagger}=\begin{cases}
a_{-k-\pi} & -\pi<k<0\\
a_{k}^{\dagger} & 0\le k\le\pi\,,\end{cases}\label{eq:the_map}\end{equation}
Notice that the particle hole transformation does not involve neither
$k=0$ nor $k=\pi$. In fact, for these two momenta, the dimer model
is given by $2(d_{0}^{\dagger}d_{0}-d_{\pi}^{\dagger}d_{\pi})$ and
the anisotropic one by $2(a_{0}^{\dagger}a_{0}-a_{\pi}^{\dagger}a_{\pi})$.
The same mapping Eq.~(\ref{eq:the_map}) transforms $\tilde{H}_{d}^{\epsilon}$
into $\tilde{H}_{XY}^{\epsilon}$ also in the case of ABC where the
momenta satisfy $k=\pi/L\left(2n-1\right)$, $n=-L/2+1,\ldots,L/2$. 

The mapping Eq.~(\ref{eq:the_map}) can be written in a compact form
as $d_{k}^{\dagger}=f_{+}\left(k\right)a_{k}^{\dagger}+f_{-}\left(k\right)a_{-k-\pi}$
with the help of two auxiliary functions $f_{\pm}\left(k\right):=\theta[\pm\sin\left(k\right)]\pm\delta_{\sin\left(k\right),0}/2$,
where $\theta$ is the Heaviside function with the convention $\theta\left(0\right)=1/2$. 

Thanks to Eq.~(\ref{eq:the_map}) the equivalence between (fermionic)
dimer and anisotropic models can be generalized. In fact, the mapping
transforms an $r$-nearest neighbor hopping term into itself, provided
$r$ is odd. Instead, an alternating hopping of the form $\sum_{i}\left(-1\right)^{i}d_{i}^{\dagger}d_{i+r}+\hc$
becomes $\sum_{i}(a_{i}^{\dagger}a_{i+r}^{\dagger}+a_{i+r}a_{i})$,
again for $r$ odd. When $r$ is even the mapping introduces non-analiticities
in Fourier space and correspondingly the transformed model becomes
long-ranged in real space. These findings can also be obtained directly
in real space Fourier transforming back Eq.~(\ref{eq:the_map}):\begin{equation}
d_{m}^{\dagger}=\sum_{x}\left[\hat{f}_{+}\left(m-x\right)a_{x}^{\dagger}+(-1)^{x}\hat{f}_{-}\left(m-x\right)a_{x}\right],\label{eq:map real space}\end{equation}
with the definition $\hat{f}_{\pm}\left(x\right)=L^{-1}\sum_{k}e^{-ikx}f_{\pm}\left(k\right)$.
Writing simply $\hat{f}_{\pm}$ in place of the matrix $(\hat{f}_{\pm})_{i,j}:=\hat{f}\left(i-j\right)$
the following relations hold: $\hat{f}_{\pm}\hat{f}_{\pm}=\hat{f}_{\pm}$,
and $\hat{f}_{+}\hat{f}_{-}=\hat{f}_{-}\hat{f}_{+}=0$.

\paragraph*{Proof of Theorem 1 }

The first step is to map the spin models Eqs.~(\ref{eq:H_d-spin})
and (\ref{eq:H_XY-spin}) to fermionic models via the JW transformation.
In terms of ladder operator $\sigma_{i}^{\pm}=\left(\sigma_{i}^{x}\pm i\sigma^{y}\right)/2$,
the JW is given by $\sigma_{i}^{+}=c_{i}^{\dagger}e^{i\pi\sum_{j=1}^{i-1}c_{j}^{\dagger}c_{j}}$
(this in turn implies $\sigma_{i}^{-}=c_{i}e^{-i\pi\sum_{j=1}^{i-1}c_{j}^{\dagger}c_{j}}$,
$\sigma_{i}^{z}=2c_{i}^{\dagger}c_{i}-\1$). The dimer and anisotropic
boundary terms become respectively \begin{align*}
H_{d}^{\eta_{d}}\to & -\eta_{d}\left(1+\gamma\left(-1\right)^{L}\right)\left[d_{L}^{\dagger}d_{1}+d_{1}^{\dagger}d_{L}\right]e^{i\pi N_{d}}\\
H_{XY}^{\eta_{XY}}\to & -\eta_{a}\left[a_{L}^{\dagger}a_{1}+\gamma a_{L}^{\dagger}a_{1}^{\dagger}\right]e^{i\pi N_{a}}+\hc,\end{align*}
where $N_{d(a)}$ is the total number operator for the $d\,(a)$ fermions
and $\eta_{d(XY)}$ specifies the spin BC for the dimer and XY model.
For OBC, $\eta_{d}=\eta_{XY}=0$, we can directly apply the result
of Theorem \ref{thm:fermi} and deduce that also the spin models are
unitarily equivalent for $L$ odd. To study the remaining cases we
first need to compute $\exp\left(i\pi N_{d}\right)$ under the action
of the mapping Eq.~(\ref{eq:the_map}). Writing the number operator
in Fourier space we get $N_{d}=\sum_{0\le k\le\pi}a_{k}^{\dagger}a_{k}+\sum_{-\pi<k<0}a_{k}a_{k}^{\dagger}$.
The sum over negative momenta contains a different number of terms
depending on the boundary conditions. For PBC the sum contains $L/2-1$
terms while for ABC it contains $L/2$ terms. Calling $N_{a}^{+}\equiv\sum_{0\le k\le\pi}a_{k}^{\dagger}a_{k}$
and $N_{a}^{-}\equiv\sum_{-\pi<k<0}a_{k}^{\dagger}a_{k}$ we can write
compactly $N_{d}=N_{a}^{+}-N_{a}^{-}+L/2-\left(1+\epsilon\right)/2$,
where $\epsilon=\pm1$ defines the boundary conditions of the fermions.
Since $N_{a}^{\pm}$ are integers, under the action of the mapping
we obtain $\exp\left(i\pi N_{d}\right)=-\epsilon\left(-1\right)^{L/2}\exp\left(i\pi N_{a}\right)$.
Let us now consider the spin models $H_{d}^{\eta_{d}}$ ($H_{XY}^{\eta_{XY}}$)
in the parity sector $\sigma_{d}$ ($\sigma_{XY}$). Thanks to the
JW transformation, for $L$ even, the parity operator is $P=e^{i\pi N_{d(a)}}$
and so in each sector $e^{i\pi N_{d(a)}}=\sigma_{d(XY)}$. This means,
that in the parity sector $\sigma_{d}$ the spin model with BC $\eta_{d}$
has boundary conditions $-\eta_{d}\sigma_{d}$ in the fermions. The
same clearly holds for the anisotropic XY model. The equivalence of
the fermionic models (Theorem \ref{thm:fermi}) holds when they have
the same BCs, that we denote with $\epsilon$. So, we arrive at the
relation $\eta_{d}\sigma_{d}=\eta_{a}\sigma_{a}=-\epsilon$. Now we
use the mapping of $\exp\left(i\pi N_{d}\right)=\sigma_{d}$, obtaining
$\sigma_{d}=-\epsilon\left(-1\right)^{L/2}\sigma_{XY}$. Solving these
last two equations, we finally obtain the parity sectors and boundary
conditions, under which the equivalence of the spin models apply:
$\sigma_{XY}=\eta_{d}\left(-1\right)^{L/2}$ and $\eta_{XY}=\sigma_{d}\left(-1\right)^{L/2}$.
$\square$

In the above proof we have partly seen what happens to the conserved
quantity $N_{d}$ after the action of the mapping. The precise form
also depends on the boundary conditions $\epsilon$. In Fourier space
we can write $N_{d}=\sum_{k}\left[f_{+}\left(k\right)-f_{-}\left(k\right)\right]a_{k}^{\dagger}a_{k}+L/2-\left(1+\epsilon\right)/2$.
Since the functions $f_{\pm}\left(k\right)$ are not analytic, and
using $f_{+}\left(k\right)+f_{-}\left(k\right)=1$, $\forall k$,
the number operator becomes non local in real space. For example its
explicit form for PBC ($\epsilon=1$) is\begin{equation}
N_{d}=\frac{2}{L}N_{a}+\frac{L}{2}-1+\sum_{x\neq y}\left(i\right)^{x-y}\frac{2\sin\left[\left(x-y\right)\left(\frac{\pi}{L}+\frac{\pi}{2}\right)\right]}{L\sin\left[\left(x-y\right)\frac{\pi}{L}\right]}a_{x}^{\dagger}a_{y}.\label{eq:N_d}\end{equation}
A corollary of our proof is that such operator commutes with the anisotropic
Hamiltonian Eq.~(\ref{eq:H_XY-Fourier}) ($\epsilon=1$) and its
spectrum is made of integers from zero to $L$.

\paragraph*{Majorana fermions}

The possibility of mapping the dimer model into an anisotropic one
is not restricted to the mapping Eq.~(\ref{eq:the_map}). Another
mapping is obtained directly in real space by introducing Majorana
fermions $\zeta_{\alpha}(j)$, $\alpha=1,2$, \begin{eqnarray*}
\left(\begin{array}{c}
\zeta_{1}(j)\\
\zeta_{2}(j)\end{array}\right) & = & \frac{1}{\sqrt{2}}\left(\begin{array}{cc}
1 & 1\\
-i & i\end{array}\right)\left(\begin{array}{c}
a_{j}^{\dagger}\\
a_{j}\end{array}\right),\end{eqnarray*}
that satisfy the commutation relations $\left\{ \zeta_{\alpha}(j),\zeta_{\beta}(j')\right\} =\delta_{\alpha\beta}\delta_{jj'}$.
This way, it is possible to show that each model gets transformed
onto two separate Ising chains in transverse field each consisting
of $L/2$ sites \cite{igloi00}. Then, assuming $L$ even and PBC
or ABC, the two pairs of Ising chains are made identical by translating
by one site one of the two chains obtained from $H_{XY}$. The composition
of all these steps yields to the following mapping \begin{equation}
d_{j}^{\dagger}=\frac{1}{2}\left[ia_{j+1}^{\dagger}+a_{j}^{\dagger}-(-1)^{j}(ia_{j+1}+a_{j})\right].\label{eq:map real space 2}\end{equation}
The transformation above has the advantage of being local in real
space and much simpler than Eq.~(\ref{eq:map real space}). By using
Eq.~(\ref{eq:map real space 2}) one can reproduce the results of
Theorem \ref{thm:fermi} for PBC or ABC. However Eq.~(\ref{eq:map real space 2})
is more powerful in view of its applications to more general local
Fermi models. Using the mapping Eq.~(\ref{eq:map real space 2})
the {}``disordered'' tight binding model \begin{equation}
H_{d}=\sum_{j=1}^{L}J_{j}d_{j}^{\dagger}d_{j+1}+\hc\label{eq:H_d-gen}\end{equation}
with arbitrary hopping rate $J_{j}$ can be mapped onto the generalized
anisotropic model \begin{equation}
H_{XY}=\sum_{j=1}^{L}\left[J_{j}^{(+)}a_{j}^{\dagger}a_{j+1}+J_{j}^{(-)}a_{j}^{\dagger}a_{j+1}^{\dagger}\right]+\hc,\label{eq:H_XY-gen}\end{equation}
with $J_{j}^{(\pm)}=(\pm)^{j}(J_{j}\pm J_{j-1})/2$ and $J_{j}$ must
always be considered periodic i.e.~$J_{L+i}=J_{i}$. This mapping
can be further generalized by adding a uniform and staggered chemical
potential. After applying the transformation Eq.~(\ref{eq:map real space 2})
such terms become\begin{multline*}
\sum_{j=1}^{L}\left[\mu+\mu_{\mathrm{st}}\left(-1\right)^{j}\right]d_{j}^{\dagger}d_{j}-\mu\frac{L}{2}=\\
-\frac{i}{2}\sum_{j=1}^{L}\left[\mu+\mu_{\mathrm{st}}\left(-1\right)^{j}\right]\left[a_{j}^{\dagger}a_{j+1}-\left(-1\right)^{j}a_{j}^{\dagger}a_{j+1}^{\dagger}\right]+\hc.\end{multline*}
The equivalence between the generalized models Eqs.~(\ref{eq:H_d-gen})
and (\ref{eq:H_XY-gen}) has potential applications in the study of
disordered systems. To obtain results on the random version of the
anisotropic model Eq.~(\ref{eq:H_XY-gen}), it can be favorable to
simulate Hamiltonian Eq.~(\ref{eq:H_d-gen}) which conserves the
number of excitations. Moreover, through the JW transformation, apart
from a possible border term depending on the BCs, the equivalence
between Fermi models can be extended to their spin counterpart. In
this way a random XY model can be mapped into a random XX model.

\paragraph*{Continuum limit}

The mappings that we have analyzed so far admit a simple interpretation
in the continuum limit. To this end we expand the fermionic fields
into chiral components $\psi(x)=e^{ik_{F}x}R(x)+e^{-ik_{F}x}L(x).$
For $\gamma=0$ the two models merge in free massless fermions: $H_{0}\equiv\sum_{j}a_{j}^{\dagger}a_{j+1}+h.c.$
where the band is half filled, so $k_{F}=\pi/2$. In the continuum
limit, we get \cite{giamarchi_book,nersesyan_book}\begin{equation}
H_{0}=i\int_{0}^{L}dx\left[\,:\negthinspace R^{\dagger}(x)\partial_{x}R(x)-L^{\dagger}(x)\partial_{x}L(x)\negthinspace:\right]\label{eq:free Dirac}\end{equation}
while the mass-generating terms in $\gamma$, $\mathcal{O}_{XY}=a_{j}^{\dagger}a_{j+1}^{\dagger}+h.c.$
and $\mathcal{O}_{d}=(-1)^{j}a_{j}^{\dagger}a_{j+1}+h.c.$ become\begin{eqnarray*}
\mathcal{O}_{XY} & = & i\,:\negthinspace L^{\dagger}(x)R^{\dagger}(x)-R(x)L(x)\negthinspace:\\
\mathcal{O}_{d} & = & i\,:\negthinspace L^{\dagger}(x)R(x)-R^{\dagger}(x)L(x)\negthinspace\colon.\end{eqnarray*}
From these expressions, we see directly that the terms multiplied
by $\gamma$ in $H_{d}$ and $H_{XY}$ are transformed into each other
by particle-hole exchange (and a minus sign) on the left movers, $L\to-L^{\dagger}$,
which is reminiscent of the discrete mapping Eq.~(\ref{eq:the_map})
where the particle hole transformation was also applied only for negative
momenta. 

Translating into bosonic language, it is known that the model Eq.~(\ref{eq:free Dirac})
is equivalent to the Gaussian model\begin{equation}
H_{0}=\frac{1}{2}\int dx\left\{ [\partial_{x}\Theta(x)]^{2}+[\partial_{x}\Phi(x)]^{2}\right\} .\label{eq:Gaussian}\end{equation}
The fields $\Phi$ and $\Theta$ are bosonic and reciprocally dual:
$\partial_{x}\Phi=\partial_{\tau}\Theta$ and $\partial_{\tau}\Phi=\partial_{x}\Theta$.
A nonvanishing value of $\gamma$ has the effect of transforming Eq.~(\ref{eq:Gaussian})
in the sine-Gordon model by adding a relevant (in the renormalization
group sense) term $\mathcal{O}_{XY}=\,:\negthickspace\sin(\sqrt{4\pi}\Theta(x))\negthickspace:$
or $\mathcal{O}_{d}=\,:\negthickspace\sin(\sqrt{4\pi}\Phi(x))\negthickspace:$,
respectively in the XY or in the dimer case. Hence, in the bosonic
language, the dimer$\leftrightarrow$XY mapping simply acts by swapping
$\Phi\leftrightarrow\Theta$. It is interesting to observe that a
direct consequence of the mapping is the interchange between density
and current density, as it can be readily inferred by their expressions\begin{eqnarray*}
\rho(x) & = & \::\negmedspace R^{\dagger}(x)R(x)+L^{\dagger}(x)L(x)\colon=-\frac{1}{\sqrt{\pi}}\partial_{x}\Phi(x)\\
j(x) & = & \::\negmedspace R^{\dagger}(x)R(x)-L^{\dagger}(x)L(x)\colon=\frac{1}{\sqrt{\pi}}\partial_{x}\Theta(x).\end{eqnarray*}
Integrating these densities over the space we obtain two quantum numbers:
the total number and the current. In particular, the total number
and current are directly related to the two winding numbers $m,n\in\mathbb{Z}$
of respectively $\Theta$ and $\Phi$. Such integers ( which determine
the scaling dimensions of the primary operators in the Gaussian model)
are both good quantum numbers for $\gamma=0$. For $\gamma\neq0$,
the breaking of translational symmetry in the dimer chain invalidates
the conservation of the current, but maintains the particle number
conservation. In the XY model the situation is just reversed: the
particle number is no more conserved, due to the pair creation-destruction
terms, while the current keeps being a good quantum number.

\paragraph*{Higher dimensions}

The mapping described in Eq.~(\ref{eq:the_map}) can be easily generalized
to $D$-dimension. In an hypercubic $D$-dimensional lattice the anisotropic
model reads\begin{equation}
\tilde{H}_{XY}=\sum_{i=1}^{D}\sum_{\mathbf{x}}\left(a_{\mathbf{x}}^{\dagger}a_{\mathbf{x}+\mathbf{e}_{i}}+\gamma a_{\mathbf{x}}^{\dagger}a_{\mathbf{x}+\mathbf{e}_{i}}^{\dagger}\right)+\hc,\label{eq:H_XY-D}\end{equation}
where $\mathbf{x}=(x_{1},\dots,x_{D})$ and $\mathbf{e}_{i}$ is the
unit vector along the $i$-th direction. We do not specify BCs here,
to fix ideas we can take PBC on a bipartite lattice. After Fourier
transforming one realizes that the BZ is contained in $[-\pi,\pi]^{D}$
. Now, let us divide the BZ in two regions according to the sign of
the first moment $k_{1}$: $A=\left\{ \boldsymbol{k}\in BZ,\,:k_{1}\in[0,\pi]\right\} $
and $B=\left\{ \boldsymbol{k}\in BZ,\,:k_{1}\in\left(-\pi,0\right)\right\} $.
The canonical transformation $a_{\mathbf{k}}^{\dagger}=d_{\mathbf{k}}^{\dagger}$
for $\mathbf{k}\in A$ and $a_{\mathbf{k}}^{\dagger}=d_{-\mathbf{k}-\vec{\pi}}$
for $\mathbf{k}\in B$ with $\vec{\pi}=\left(\pi,\pi,\ldots,\pi\right)$,
generalizes the one-dimensional version Eq.~(\ref{eq:the_map}).
This mapping transforms the Hamiltonian Eq.~(\ref{eq:H_XY-D}) into
the $D$-dimensional dimer model:\begin{eqnarray*}
\tilde{H}_{XY} & = & \sum_{i=1}^{D}\sum_{\mathbf{k}}\left[2\cos(\mathbf{k}\cdot\mathbf{e}_{i})d_{\mathbf{k}}^{\dagger}d_{\mathbf{k}}\right.\\
 &  & \left.+\left(i\gamma\sin(\mathbf{k}\cdot\mathbf{e}_{i})d_{\mathbf{k}}^{\dagger}d_{\mathbf{k}+\vec{\pi}}+\hc\right)\right]\\
 & = & \sum_{i=1}^{D}\sum_{\mathbf{x}}\left(1+\gamma\left(-1\right)^{\left|\boldsymbol{x}\right|}\right)d_{\mathbf{x}}^{\dagger}d_{\mathbf{x}+\mathbf{e}_{i}}+\hc\,,\end{eqnarray*}
where the modulation factor is given by $\left(-1\right)^{\left|\boldsymbol{x}\right|}=\exp\left(i\vec{\pi}\cdot\boldsymbol{x}\right)$.

\paragraph*{Conclusions}

In this paper we have analyzed two common spin models (XY and dimerized
XX) and showed that they are unitary equivalent apart from at most
a border term. By explicitly providing the unitary transformation
we have been able to generalize the equivalence in many ways. For
example the fully disordered (with site dependent couplings) XY chain
can be mapped onto a disordered XX chain. Considering the fermionic
counterpart we have also shown that generally a dimerized, $r$-nearest
neighbor, hopping term, is mapped onto an $r$-nearest neighbor pair
creation term. In one dimension our mappings have a simple interpretation
in the continuum limit in terms of bosonic fields. Similar considerations
can also be extended to higher dimensions. 

Mapping XY models onto XX ones can be useful in view of numerical
simulations of disordered models or non-integrable extensions. This
is due to the explicit particle number conservation of the XX models
which makes them easier to treat numerically. A by-product of our
analysis is that particle number symmetry is also present in the XY
models although in a hidden fashion.

\paragraph*{Acknowledgments }

We are grateful to T. Giamarchi for inspiring us the picture in the
continuum. We also thank J.I.~Cirac, and Z.~Zimboras for interesting
discussions. We have been supported by the EU-STREP Projects HIP (grant
no. 221889) and COQUIT (grant no. 233747).

\end{document}